\documentclass[aps,axodraw,twocolumn,prl]{revtex4}
\def\bea{\begin{eqnarray}}
\def\eea{\end{eqnarray}}
\def\a{\alpha}
\def\D{\Delta}
\def\f{\frac}
\def\l{\lambda}

\def\p{\prime}
\def\S{\Sigma}
\def\nn{\nonumber}
\tighten \preprint{SNUTP 04-015}
\begin{document}
\title{\Large\bf 
Electroweak Symmetry Breaking from SUSY Breaking\\
with Bosonic See-Saw Mechanism}
\author{
Hyung Do Kim
\footnote{hdkim@phya.snu.ac.kr}}
\address{
School of Physics, Seoul National University, Seoul
151-747, Korea} 

\begin{abstract}
We introduce the idea of {\it bosonic see-saw mechanism}
in analogy with the see-saw mechanism.
Bosonic see-saw is a new symmetry breaking mechanism
and we apply it to explain electroweak symmetry breaking 
as an inevitable consequence of supersymmetry breaking.
The breaking of electroweak symmetry occurs at tree level
once supersymmetry is broken. 
Absence of color/charge breaking
in this model is related to doublet-triplet splitting
in grand unified theory.
An extension of MSSM with a weak triplet
shows very interesting results especially when $\mu =0$.
It provides the most natural understanding
of why we have only electroweak symmetry breaking
rather than having color/charge breaking.
In the limit $\mu=0$, the model predicts very light chargino mass,
104 GeV while Higgs is heavy, 130 GeV.
\end{abstract}
 
\maketitle
\thispagestyle{empty}

The standard model(SM) has a beautiful structure of explaining
all the matters and forces except gravity in terms of quark/lepton(s)
and gauge interactions.
All the quarks and leptons are massless
as long as electroweak symmetry is unbroken,
and they can get mass from Yukawa interactions 
only after electroweak symmetry breaking.
Within the framework of the standard model,
Higgs potential can be arbitrary and we choose
the sign of the coefficient of quadratic (quartic) term
to be negative (positive) such that the Higgs potential
has a desired mexican hat shape.
It would not be easy to understand why the quadratic term
has a negative sign while the quartic term is positive
within the standard model.

The SM is just regarded as a low energy effective theory
of some extended one and gauge hierarchy problem suggests
a modification of the standard model at TeV scale.
Supersymmetry(SUSY) \cite{Nilles:1983ge,Haber:1984rc} 
is one of the most promising candates for it.
In supersymmetric extensions of the standard model,
we can get a better understanding of the electroweak symmetry breaking.
First of all, the Higgs potential is no longer arbitrary
and should be a sum of supersymmetric F/D terms and
soft supersymmetry breaking terms.
The quartic term is calculated from gauge couplings and
is positive definite. The quadratic term (soft terms) is
a sum of supersymmetric mass term ($\mu$ term)
and soft supersymmetry breaking terms which are calculable
in certain mediation mechanism of supersymmetry breaking.
In the minimal supersymmetric standard model(MSSM)
with gauge mediated SUSY breaking \cite{Dine:1993yw,Dine:1994vc,Dine:1995ag},
a radiative correction by large top Yukawa coupling
gives negative Higgs mass squared.
If $\mu$ were zero, the above picture might have been beautiful
and could be considered as a possible explanation of the electroweak
symmetry breaking.
However, in reality, the electroweak symmetry breaking
is a surprising cancellation of $\mu^2 \sim (500 \ {\rm GeV} )^2$
and the Higgs soft scalar mass squared $m_{H_u}^2$ \cite{Chang:2004db}.
If $\mu$ were slightly larger, we would never have the electroweak
symmetry breaking. And if $\mu$ were slightly smaller,
the Higgs would develop its vacuum expectation value (VEV)
exponentially larger than the weak scale.
Thus it is desirable to consider 
models in which the weak scale electroweak symmetry breaking can be explained
for a broad range of parameters.

In this paper we first introduce a simple idea called
'bosonic see-saw mechanism'.
We apply 'bosonic see-saw mechanism'
to explain the electroweak symmetry breaking.
When down type Higgs couples to the extra vector-like pair of Higgs,
we can soften the little hierarchy problem.
Finally we consider a model in which 
the electroweak symmetry is closely tied up to the supersymmetry

Let us briefly discuss the conventional see-saw mechanism explaining
the lightness of neutrino masses \cite{Gell-Mann:1980vs,Yanagida:1979as}.
For $\nu$, the left-handed neutrino which is an $SU(2)_L$ doublet,
and $N$ a singlet, the possible interactions are
\bea
{\cal L_{\nu}} & = & - M N N + \l_{\nu} H L N + {\rm h.c.},
\eea
where $H = (H^+ , H^0)$ is the Higgs doublet and $L = (\nu, l^-)$.
After the electroweak symmetry breaking, it becomes
\bea M_{\nu} \left(
\begin{array}{c}
  \nu \\
  N \\
\end{array}
\right) & = &
\frac{1}{2} \left(
\begin{array}{cc}
  0 & m_D \\
  m_D & M \\
\end{array}
\right)
\left(
\begin{array}{c}
  \nu \\
  N \\
\end{array}
\right), \eea
where $m_D = \l_{\nu} \langle H^0 \rangle$.
The lightest neutrino mass for $m_D \ll M$ is then
\bea
m_{\nu} & = & - \frac{m_D^2}{M}.
\eea
Note the sign of the lighter eigenvalue.
As the determinant of the matrix is negative definite (-$m_D^2$)
and the heavier one is nearly $M$, 
the lighter eigenvalue is negative definite.
The result is valid as long as $m_D \ll M$.
We can make the mass term to be positive definite
by the field redefinition of neutrinos.
Therefore, this observation is not important for neutrinos (fermions)
but it will turn out to be very important for later consideration.

Bosonic see-saw mechanism works for bosons instead of fermions (neutrinos).
Although the mechanism works for any scalar fields (superfields),
here we take Higgs as an example for a clear illustration.
Supersymmetric extension of the standard model requires 
two Higgs chiral superfields $H_u$ and $H_d$.
Suppose there is an additional massive pair $H_u^{\prime}$
and $H_d^{\prime}$, the electroweak doublets with the opposite
hypercharge.
Let $X = \cdots + F_X \theta^2$ be a superfield 
representing supersymmetry breaking
$F = \langle F_X \rangle \neq 0$.
For the superpotential
\bea
W & = & \l_1 X H_u^{\prime} H_d + M H_u^{\prime} H_d^{\prime} ,
\eea
the scalar mass squared matrix for $H_d, H_u^{\prime *}$ is
\bea
{\hat{\cal M}}^2 & = & \left( \begin{array}{cc} 0 & \l_1^* F^* \\
\l_1 F & |M|^2 \\
\end{array}
\right). \eea

When $\sqrt{F} \ll M$, the lightest scalar mass squared becomes
negative definite,
\bea 
m_{H_d}^2 & = & - \left|\f{\l_1 F}{M}\right|^2. \nn
\eea
Whenever $F \neq 0$, the mass squared is negative
and we end up with symmetry breaking.
Therefore, at tree level,
we obtain the electroweak symmetry breaking as a consequence of
supersymmetry breaking. We can do the same thing to $H_u$
instead of $H_d$.
Note that the sign here is physical
as the matrix is for scalar mass squared. It is called
'bosonic see-saw mechanism' as it is opposed to
usual see-saw mechanism which works for the fermions.

The bosonic see-saw mechanism shows a similarity 
to the (fermionic) see-saw mechanism.
\begin{itemize}
\item There are heavy states.
(heavy Higgs v.s. $N$)

\item There are interactions between heavy and massless states.
($H_u^{\prime}$ and $H_d$ v.s. $\nu$ and $N$)

\item Off-diagonal elements are generated if fields get VEVs.
($X \rightarrow \langle F_X \rangle \neq 0$ v.s.
$H \rightarrow  \langle H^0 \rangle \neq 0$)
\end{itemize}

The crucial difference is the negative sign
of bosonic see-saw mechanism which can not be eliminated 
by rephasing scalar fields.
In general, $\langle X \rangle \neq 0$ and we can redefine fields
and couplings such that $\tilde{X} = X - \langle X \rangle$
does not have a scalar VEV ($\langle \tilde{X} \rangle =0)$.
Then we obtain more general superpotential,
\bea
W & = & \l_1 X H_u^{\prime} H_d 
+\l_2 X H_u^{\prime} H_d^{\prime} 
+ M H_u^{\prime} H_d^{\prime}.
\eea
Now Yukawa couplings are
\bea
W & = & \l_u H_u Q u^c + \l_d H_d Q d^c + \l_u^\p H_u^\p Q u^c.
\eea

From now on, we focus on the application of bosonic see-saw mechanism
to the electroweak symmetry breaking.
As we have two Higgs fields $H_u$ and $H_d$ in MSSM,
there are three possibilites. First, $H_u$ couples to heavy Higgs.
Second, $H_d$ couples to it. Finally, both of them couple to it.
When there is no radiative correction, the first option looks
the most natural. However, we know that top Yukawa gives large radiative
corrections and the second option is the best.
If both of them couple to heavy Higgs and $X$, we can not make them light
and the third option does not work.
Thus we consider only the second possibility in this paper.

As $H_u^\p$ and $H_d^\p$ couple to $X$ directly,
they are the messengers of SUSY breaking and $M$ is the messenger scale.
We can calculate soft terms mediated by gauge interactions.
We also assume that there is a pair of color triplet Higgs fields
which complete the messenger fields into $SU(5)$ multiplets.
The soft terms from gauge mediation are positive definite 
\cite{Giudice:1998bp}.
\bea
m^2_{\Phi} & = & \sum_i 2 c_i \left( \frac{\alpha}{4\pi} \right)^2 
\Lambda^2.
\eea
where $\Lambda = |\l_2 F/M|$ and 
$c_i$ is the quadratic Casimir of $i$-th gauge group.
Note that $\l_1 \sim 10^{-2} \l_2$ is required to have
$\Lambda \sim 10$ TeV while $m_{H_d}^2 \sim 100$ GeV.
Till now the only difference with the usual MSSM is the tree level
contribution from bosonic see-saw which is negative definite.

The most interesting consequence comes with the addtion of
the electroweak triplet $\S$. Let us explain why we need $\S$
and how it brings an interesting result.

In MSSM, the nice mechanism of radiative electroweak symmetry breaking
is spoiled by large $\mu$ term.
Large $\mu$ term  in MSSM is due to the fact that
we have not seen Higgs yet.
In MSSM, the quartic couplings are given by gauge couplings
and Higgs mass is predicted to be light.
$m_H^2 > 114$ GeV requires a large radiative correction 
and it is possible only with heavy stop.
If stop is heavy, radiative corrections are too large
and the electroweak symmetry breaking becomes large
unless large $\mu$ term cancels it.
The lightness of Higgs mass in MSSM is mainly due to
the small quartic terms from gauge interactions
and it can be relaxed if there are additional quartic
couplings in the theory in addition to the usual D-term. 
Thus we consider the modification of MSSM to give the addtional
quartic terms.

The most transparent application of the bosonic see-saw mechanism
comes out if $\mu = 0$.
However, the limit $\mu = 0$ in MSSM poses several problems
\cite{Kim:1983dt}.
\begin{itemize}
\item Peccei-Quinn(PQ) symmetry and R symmetry

As Higgs fields carry PQ and R charge
and the symmetry is exact in the limit $\mu =0$,
once they get VEVs, there appears a massless Goldstone boson
which is in confict with experiments.

\item Electroweak symmetry breaking

If $H_u$ gets a VEV from negative $m_{H_u}^2$,
$H_d$ gets its VEV through $B\mu$ term.
Therefore, if $\mu = 0$ ($B\mu=0$),
the down-type quarks and charged leptons can not get their masses.

\item Chargino mass

If $\mu =0$, higgsino can get their mass only by the electroweak
symmetry breaking and the lightest chargino mass is always
lighter than $M_W$ which can not be compatible with 
the current bound on the lightest chargino mass, 104 GeV.
\end{itemize}

These problems can be solved if extra fields are introduced.
In NMSSM, an extra singlet replaces $\mu$ term.
The singlet gets a VEV and it generates $\mu$ term effectively.
Then the electroweak symmetry breaking is a fine tuning just as in MSSM.
An alternative way is to introduce an extra weak triplet $\Sigma$
with no hypercharge.

Let us consider the most interesting limit $\mu=0$ ($\mu$-less SSM).
We can forbid $\mu$ term by a discrete symmetry, so called 'U parity',
which is a $Z_2$ subgroup of Peccei-Quinn symmetry.
Under the U parity,
\bea
(H_u,u^c,\S) & \rightarrow & - (H_u,u^c,\S). \nn 
\eea

The most general superpotential consistent with the U parity is
\cite{Espinosa:1991wt}
\bea
W & = & \frac{M_{\Sigma}}{2} {\rm tr} \Sigma^2 
+ \l_{\Sigma_1} H_u \Sigma H_d
+ \l_{\Sigma_2} H_u \Sigma H_d^{\prime}.
\eea
These terms are enough to break PQ and R symmetry. At the same time
chargino mass can be heavier than $M_Z$ as we have new sources for it.

Soft supersymmetry breaking terms are
\bea
V_{\rm soft} & = & m_{H_u}^2 |H_u|^2 
+m_{H_d}^2 |H_d|^2 
+m_{\Sigma}^2 {\rm tr} \Sigma^{\dagger} \Sigma \nn \\
&&
+ A \l_{\Sigma_1} H_u \Sigma H_d 
+ B M_{\Sigma} {\rm tr} \Sigma^2 + {\rm h.c.}.
\eea

The neutral component of $\Sigma$ gets a VEV once $H_u$ and $H_d$ get VEVs
\cite{Espinosa:1991wt},
\bea
v_{\Sigma} & = & \frac{\l_{\S_1} \frac{M_{\Sigma}}{2} v^2}{
m_{\Sigma}^2 + M_{\Sigma}^2 + B_{\Sigma} M_{\Sigma} 
+ \frac{1}{2} {\l_{\S_1}}^2 v^2}.
\eea
$v_{\Sigma} < 9$ GeV is obtained if $m_{\Sigma}$ is larger than
the electroweak scale.

Let us go back to the calculation of soft terms.
As our messenger fields have direct couplings with matter/Higgs fields,
there are additional contributions.
The Yukawa mediated ones are calculated using the formalism
of analytic continuation into superspace
\cite{Giudice:1997ni,Arkani-Hamed:1998kj,Chacko:2001km},
\bea
\D m_{H_u}^2 & = & \left[ \f{3 \a_{\S_2}^2 }{4\pi^2}
+ \f{\a_{\S_2} \a_{\l_2} }{4\pi^2}\right] 
\Lambda^2 \nn \\
\D m_{H_d}^2 & = & \left[ -\f{\a_{\S_1} \a_{\S_2} }{2\pi^2}\right] 
\Lambda^2 \nn \\
\D m_{Q,u^c}^2 & = & \left[ -\f{\a_{t} \a_{\S_2} }{8\pi^2}\right] 
\Lambda^2 \nn \\
\D m_{\S}^2 & = & \left[ \f{3 \a_t \a_{\S_2} }{4\pi^2}
+ \f{3 \a_{\S_2}^2 }{4\pi^2}
+ \f{ \a_{\S_2} \a_{\l_2} }{4\pi^2}
- \f{5 \a \a_{\S_2} }{4\pi^2}
\right] 
\Lambda^2, \nn
\eea
where $\a = \f{g^2}{4\pi}$ is the $SU(2)$ gauge coupling
and $\a_{f} = \f{f^2}{4\pi}$ are similarly defined Yukawa couplings
for $f = \l_1, \l_2, \l_{\S_1}, \l_{\S_2}, \l_t$.
We assume $\l_t^\p \ll 1$ and neglects its contribution.
Other Yukawa couplings are also neglected as they are small.
The effects are summarized as follows.
\begin{itemize}
\item $H_d$ :
Soft scalar mass squared is negative at the tree level 
from the bosonic see-saw mechanism.
There are threshold corrections from gauge and Yukawa interactions
and the sign is opposite. If Yukawa and gauge couplings are of similar size,
the threshold corrections at the messenger scale cancel with each other.
Therefore, negative mass squared at the tree level dominates.

\item $H_u$ :
Threshold corrections at the messenger scale are positive for both
gauge and Yukawa contributions.
We have slightly larger $m_{H_u}^2$ compared to the MSSM with gauge mediation.
We should also consider negativ one loop correction 
from messenger scale to the weak scale
$-\f{3}{4\pi^2} m_{\tilde{t}}^2 \log \f{M}{m_{\tilde{t}}}$.

\item $\S$ :
Threshold corrections are positive for gauge and Yukawa contributions.
Thus we get $m_\S^2$ heavier than other soft scalar masses
which is necessary to suppress the VEV of $\S$ compared to $H_u$ and $H_d$.

\item Third generation $Q, u^c$ ($\rightarrow$ stop)

Threshold correction from Yukawa mediation is negative.
We get lighter stop mass compared to the MSSM
which makes the negative contribution to $m_{H_u}^2$ smaller than usual.

\end{itemize}
The most challenging phenomenological constraint 
comes from chargino mass bound
combined to the precision data.
The chargino mass is obtained from 
Yukawa interactions (A : Higgsino-Wino-Higgs and B : Higgsino-$\psi_{\S}$-Higgs)
in the $\mu$-less theory \cite{Nelson:2002ca}.
A is the gauge coupling and B is a new Yukawa coupling $\l_{\S_1}$
that violates custodial $SU(2)$ symmetry.
The bound on the precision variable $T$  ($T < 0.6$) restricts
$\l_{\S_1}$ ($\l_{\S_1} < 0.6$) \cite{Nelson:2002ca}.
For $\l_{\S_1} \sim 1$, we obtain the lightest chargino mass to be 104 GeV
which is the bound from LEP II. The chargino masses are $(104,119,252)$ GeV
for $\l_{\S_1} = 1, (M_2, M_{\S}) = (120,150)$ GeV.
Note that if we allow nonzero $\mu$, we can satisfy the chargino mass bound
with a smaller $\l_{\S_1}$. 
More precise calculation of $T$
is needed as we deal with light spectrum (charginos are near 100 GeV).
For the neutralinos, the lower mass bound 40 GeV is easily satisfied.

The Higgs mass can be calculated if all the parameters are chosen.
As we have a new quartic couplings for the Higgs from $W = \l_1 H_u \S H_d$,
the lightest scalar Higgs mass is heavier than the one in the MSSM
and is around 120 to 130 GeV before considering the one loop correction.
Therefore, in this model we do not need to tune the parameters
to raise up the Higgs mass beyond the current bound 114 GeV.
The bosonic see-saw mechanism gives $m_{H_d}^2 < 0$
at the messenger scale and $m_{H_u}^2$ is driven to be negative
by RG running to the weak scale.
Both $m_{H_u}^2$ and $m_{H_d}^2$ are negative at the weak scale
and the minimum is at around $\tan \beta = \f{v_u}{v_d} \sim {\cal O}(1)$.
Unlike in the MSSM, the potential is not bound from below
for $m_{H_u}^2 < 0$,  $m_{H_d}^2 < 0$ as the new quartic coupling $\l_1$
prevents them from running away along D-flat direction.

In this paper we proposed a new mechanism to understand the electroweak
symmetry breaking.
As $H_d$ couples directly to the messenger of supersymmetry breaking,
the soft scalar mass squared is negative by the bosonic
see-saw mechanism when supersymmetry is broken.
The soft scalar mass squared of $H_u$ is driven to be negative
and the symmetry breaking minimum is at around 
$\tan \beta  = \f{v_u}{v_d} \sim {\cal O}(1)$.
There is a new $SU(2)_L$ triplet $\S$ which couples to $H_u$ and $H_d$.
The lightest chargino mass is predicted to be light
due to the absence of supersymmetric mass $\mu$
and lies just above the current mass bound 104 GeV.
The lightest Higgs mass is heavy as we have a new quartic coupling.
All the soft parameters appear from
gauge mediation and new Yukawa(Higgs) mediation
and they are calculable.
Gauge mediation gives positive definite soft scalar masses
which guarantees the absence of color/charge breaking minima.
Yukawa(Higgs) mediation gives negative contributions to $H_d$
and the third generation $Q$ and $u^c$ (stop)
and positive contributions to $H_u$ and $\S$.
The contributions of Yukawa(Higgs) mediation softens
the little hierarchy problem of MSSM.
As Higgs can be heavy, the fine tuning problem is no longer severe.

The setup considered here naturally arise from 
the five dimensional geometric setup \cite{Kim:2002im}.
The orbifold GUT fixes the location of gauge and Higgs fields
to be in the bulk and the distant brane is a source of supersymmetry breaking.
In this case Higgs is very special and can feel the supersymmetry breaking
directly. The massive vector-like fields introduced here is just 
the massive Kalaza-Klein towers of bulk Higgs fields.
Gaugino mediation can be considered at the same time 
as there is no symmetry preventing the couplings of supersymmetry breaking
fields with gauge sector.
In the orbifold GUT, the doublet-triplet splitting of Higgs fields
is explained by the boundary condition (or orbifold projection)
and the setup given in this paper naturally arises from
higher dimensions.
More precisely, only $H_d$ should be bulk fields as in \cite{Kim:2001at}.
The setup has been studied to understand
the top/bottom mass hierarchy without large $\tan \beta$ in \cite{Kim:2001at}.
Furthermore, the smallness of $\l_1$ compared to $\l_2$ can be explained by
the zero mode localization of $H_d$
\cite{Arkani-Hamed:1999dc,Kitano:2003cn,Kim:2004vk}.

We proposed a new idea called 'bosonic see-saw mechanism'.
Once supersymmetry is broken,
at same time it gives the VEV to the Higgs fields, i.e.,
quarks and leptons get their masses.
The mechanism works nicely even if $\mu=0$ though we need
additional weak triplet.
The chargino remains light (near 104 GeV) when $\mu=0$ 
and it is robust against radiative corrections.
Higgs is heavy (about 130 GeV before considering radiative corrections) but
the full spectrum of Higgs can be obtained only after considering
the radiative corrections and we leave the detailed calculation of it
for future work. 
The bosonic see-saw mechanism can be applied differently in other problems.

\begin{acknowledgments}
This work is supported by the ABRL Grant No. R14-2003-012-01001-0,  
the BK21 program of Ministry of Education, Korea.
\end{acknowledgments}

\end{document}